**Advancing Equity in STEM: A Critical Analysis of NSF's Division for Equity and Excellence in STEM through Theoretical Lenses**

Shaouna Shoaib Lodhi

2## Abstract

This paper critically examines the National Science Foundation's (NSF) Division of Equity for Excellence in STEM (EES) initiatives, specifically focusing on their mission to broaden participation among historically underrepresented groups in STEM fields. Grounded in the literature on equity in STEM education, the paper addresses both agreements and criticisms of the NSF's goals. While acknowledging the importance of diversity for fostering innovation and maintaining U.S. competitiveness, the analysis reveals that current policies often fall short in addressing systemic barriers that hinder equitable participation. The paper proposes key research questions aimed at understanding the impact of racial hierarchies and systemic inequities embedded within STEM education policies. Using Critical Race Theory (CRT) and Mills's Racial Contract as theoretical frameworks, the study examines how these policies perpetuate racial commodification and exclusion, ultimately reinforcing inequities. CRT is selected as the most appropriate framework to address these questions, given its focus on systemic racism and the lived experiences of marginalized students. The paper concludes with recommendations for policy reforms that go beyond access, advocating for structural changes that affirm the personhood and intellectual capabilities of underrepresented groups, thereby advancing NSF's commitment to equity and inclusion in STEM.

**Gap-Gazing Fetish: A Critical Examination of Equity in STEM Education**

Equity initiatives in STEM education often focus on identifying and closing achievement gaps between different demographic groups. This emphasis, sometimes described as the "gap-gazing fetish," refers to the fixation on comparative metrics that highlight disparities between marginalized groups and their White, middle-class counterparts (Gutiérrez, 2008). While this approach aims to bring attention to educational inequities, it inadvertently perpetuates a deficit-oriented view that frames underrepresented students as inherently lacking. By fixating on gaps, these efforts often overlook systemic issues and fail to address the complex socio-political contexts that contribute to educational inequities.

Gap-gazing tends to emphasize quantitative measures of success, such as test scores and graduation rates, without interrogating the root causes of these disparities. Scholars argue that this approach positions underrepresented groups as problems to be fixed rather than as individuals with unique strengths and experiences (Valencia, 2019; Yosso, 2005). Instead of challenging inequitable structures, gap-gazing reinforces the status quo by upholding the dominant group's standards as the ideal, thereby marginalizing the cultural capital of historically excluded groups.

In the context of STEM education, gap-gazing often translates into superficial interventions that prioritize raising the performance of marginalized students to match that of their peers, without critically examining the educational environment itself. This can manifest in the overemphasis on remedial programs, standardized test preparation, and "one-size-fits-all" solutions that fail to consider the systemic and institutional barriers that shape students' experiences (Basile & Lopez, 2015). The gap-gazing fetish, therefore, not only narrows the scope of equity work but also continues the inequities it seeks to address.

In this paper, I critically examine the equity and social justice implications of STEM education policies, specifically focusing on the NSF's "Equity for Excellence in STEM" (EES) initiative. To frame this analysis, I draw on two theoretical perspectives that underscore the racial dynamics inherent in educational policies and practices. First, I employ Mills's (1997) concept of the "racial contract" as articulated by Shah (2019), which explicates how race establishes a hierarchy of personhood, delineating who is deemed fully human (knower) and who is relegated to subpersonhood (subknower). This framework reveals how structural inequities are perpetuated through policies that appear race-neutral but often marginalize underrepresented groups. Second, I integrate Critical Race Theory (CRT), which centers race and racism as fundamental organizing principles of social structures, including education (Ladson-Billings & Tate, 1995; Yosso et al., 2009). CRT's emphasis on racial representation and critique of colorblind approaches allows for a deeper analysis of how policies like EES, while well-intentioned, may inadvertently reinforce existing disparities by failing to address the specific needs of marginalized communities. By synthesizing these theoretical frameworks, I investigate how EES policies impact STEM education at both the macro level (policy) and the micro level (classroom practices), highlighting the importance of equitable interventions that recognize and challenge systemic racism.

**Research Questions**

- RQ 1: How do STEM education policies and practices contribute to the perpetuation of racial hierarchies and inequities among underrepresented groups through the lens of Critical Race Theory (CRT)?
- RQ 2: What role does racial commodification play in shaping the experiences of underrepresented students in STEM education, and how does it affect their academic and social outcomes?

- RQ 3: How can STEM educators and institutions move beyond deficit-based narratives and adopt practices that affirm the personhood and intellectual capabilities of underrepresented students?

**Theoretical Frameworks**

In addressing the NSF's goals for diversity, equity, and inclusion within STEM education, this paper draws on two theoretical frameworks: Critical Race Theory (CRT) and the Racial Contract. Each framework offers unique insights into understanding how STEM education policies and practices impact underrepresented groups, particularly in terms of equity and social justice. By applying these frameworks, this paper critically examines how current educational policies perpetuate racial inequities and suggests approaches for more inclusive and affirming practices.

**Critical Race Theory (CRT)**

CRT is a focused framework that emphasizes the centrality of race and racism in shaping societal structures, including educational policies. Originating from legal studies, CRT challenges the normalization of racism in social institutions, revealing how policies and practices often reinforce racial hierarchies (Delgado & Stefancic, 2023).

In the context of STEM education, CRT exposes how policies designed to be "neutral" or "colorblind" often mask systemic barriers faced by students of color. It critiques ideologies like meritocracy, which overlook the historical and social inequalities affecting underrepresented groups in STEM (Solórzano & Yosso, 2001). By examining how these policies contribute to inequitable outcomes, CRT calls attention to the need for reforms that directly address the racialized barriers within STEM education.

A core component of CRT is its use of counter-narratives, which amplify the voices of marginalized students. Instead of viewing students of color through a deficit lens, CRT emphasizes their cultural wealth and intellectual potential, advocating for educational policies that affirm their personhood and capabilities (Yosso, 2005). This shift challenges deficit-based views and promotes systemic changes that focus on equity.

Overall, CRT is essential for critically analyzing STEM education policies as it highlights how these policies often perpetuate racial inequities and advocates for reforms that prioritize the experiences and strengths of underrepresented students.

**The Racial Contract**

The Racial Contract, as conceptualized by Charles Mills (1997), offers a critical framework for understanding how racial hierarchies are maintained through societal agreements. Mills argues that race is a fundamental organizing principle that positions white individuals as fully recognized persons, while people of color are relegated to a subperson status with restricted rights and recognition. This unspoken agreement perpetuates racial inequalities and shapes societal power dynamics.

In the context of STEM education, the Racial Contract provides insight into how educational policies often sustain racial disparities. For example, policies emphasizing standardized testing and advanced placement are framed as mechanisms for enhancing rigor and competitiveness in STEM. However, these policies frequently overlook systemic barriers faced by underrepresented students, such as inadequate access to quality science instruction, culturally relevant curricula, and supportive learning environments (Shah, 2019). The Racial Contract reveals how such policies, while ostensibly neutral, reinforce existing power structures by upholding the dominant group's standards and marginalizing diverse perspectives.

Additionally, the Racial Contract highlights the invisibility of whiteness in policy discourse. Educational equity is often discussed in terms of increasing access rather than addressing the deeper, structural issues that disadvantage students of color. This framework challenges the assumption of neutrality in educational policies and calls for a more explicit recognition of racial dynamics.

By exposing the hidden agreements that sustain racial hierarchies, the Racial Contract underscores the need for policies that not only enhance access but also actively dismantle systemic barriers. This perspective is crucial for advancing the NSF's goals for diversity, equity, and inclusion in STEM education, advocating for a more profound and equitable approach to policy development and implementation.

**Moving Forward with CRT over Mill's Racial Contract**

Critical Race Theory (CRT) is more suitable than Mills's Racial Contract for this paper because CRT specifically addresses race and systemic racism within institutions, including education. It focuses on how policies perpetuate racial hierarchies, critiques deficit-based narratives and centers the voices of marginalized students. CRT provides a direct lens for understanding the structural barriers that affect students of color in STEM, making it a more targeted framework for analyzing educational inequities compared to the broader societal critique offered by Mills's Racial Contract.

**Rationale for Choosing Critical Race Theory (CRT)**

While both CRT and the Racial Contract offer valuable insights into the racial dynamics of STEM education, CRT is particularly well-suited to frame the research questions of this paper. CRT's emphasis on the systemic nature of racism, its critique of neutrality and meritocracy, and its advocacy for counter-narratives align closely with the goals of examining and addressing inequities within STEM education.

CRT is instrumental in framing research questions that seek to understand how STEM education policies perpetuate racial hierarchies, the role of racial commodification, and the ways to move beyond deficit-based narratives. CRT's focus on counter-narratives allows for the amplification of marginalized voices, directly challenging the deficit perspectives that dominate STEM education. This approach is crucial for developing more equitable and inclusive policies that affirm the cultural wealth and personhood of underrepresented students.

By using CRT as the primary theoretical framework, this paper not only critiques the inequitable aspects of current STEM education policies but also advocates for transformative approaches that align with the NSF's call for diversity, equity, and inclusion. CRT provides the tools to not only identify the gaps in current policies but also envision pathways toward a more just and inclusive STEM education landscape.

**Answer to RQ 1: How do STEM education policies and practices contribute to the perpetuation of racial hierarchies and inequities among underrepresented groups through the lens of Critical Race Theory (CRT)?**

Critical Race Theory (CRT) offers a vital lens for analyzing how STEM education policies and practices contribute to the perpetuation of racial hierarchies and inequities among underrepresented groups. CRT makes race and racism central to its analysis, emphasizing that racism is deeply embedded in social structures, including education, and often goes unchallenged due to its normalization within policy and practice (Duncan, 2002). By interrogating these racialized dynamics, CRT reveals how STEM education, despite its suggested goals of inclusivity, often upholds systemic inequities.

### *The Centrality of Racism in STEM Policies*

CRT posits that racism is not an anomaly within educational policies but a fundamental and pervasive element that shapes the experiences of underrepresented groups (Ladson-Billings, 1998). In the context of STEM education, policies that emphasize standardized science testing or Advanced Placement (AP) science courses often overlook the racialized barriers that hinder students of color from accessing high-quality science education. This neglect perpetuates a system where students of color remain marginalized, aligning with CRT's assertion that education policies can covertly sustain white supremacy by failing to address the specific needs of racial minorities (Delgado & Stefancic, 2023). Such policies frequently operate under an impression of neutrality or meritocracy, if all students have equal access to resources and opportunities. This assumption obscures the deeply rooted inequities that continue to disadvantage students of color, thereby maintaining existing power structures within STEM education.

### *Racial Hierarchies and Educational Inequality*

Critical Race Theory (CRT) confronts the flawed concept of colorblindness in educational policies, emphasizing that such approaches dismiss the unique challenges faced by students of color and sustain existing racial hierarchies (Bonilla-Silva, 2015). In STEM education, this issue is evident in policies that adopt uniform strategies, such as standardized science curricula or high-stakes assessments, without acknowledging the historical and systemic inequities that affect underrepresented groups. For instance, equity-driven science initiatives often seem inclusive. Yet, they fail to address barriers like limited access to advanced science courses, experienced science educators, and culturally responsive teaching methods for students of color (Martin, 2009). Consequently, these students frequently encounter obstacles that hinder their success in advanced STEM fields, reinforcing the false perception that they are less capable than their white counterparts. This perpetuates a cycle of inequity, underscoring CRT's assertion that educational policies, while outwardly neutral, often uphold racial disparities.

### *Racial Essentialism and the Achievement Gap*

CRT's concept of racial essentialism is a practice that attributes fixed characteristics to racial groups that help explain how STEM policies contribute to the perpetuation of inequities. Educational systems often rely on deficit narratives that pathologize students of color, viewing them through a lens of racial essentialism that frames their academic struggles because of cultural or genetic deficiencies (Basile & Lopez, 2015). The focus on the achievement gap, a dominant theme in STEM education, exemplifies this perspective by framing the lower performance of students of color as a failure to meet white normative standards rather than because of systemic racism (Shah & Leonardo, 2016). This gap-gazing approach ignores the broader sociopolitical factors that create and sustain educational inequities, instead perpetuating racial hierarchies by upholding the belief that students of color are inherently less capable.

### *Commodification of Racialized Bodies in STEM*

CRT also introduces the concept of racial commodification, which critiques how the educational system exploits the presence of students of color to serve institutional diversity goals without genuinely addressing their needs (Basile & Lopez, 2015). In STEM education, this can be seen in efforts to increase minority participation rates in STEM programs without dismantling the structural barriers that hinder their success. Schools may highlight the enrollment of students of color to project an image of inclusivity while simultaneously failing to provide the resources and support necessary to ensure their academic achievement. This commodification reduces students of color to mere statistics in the pursuit of institutional goals, reinforcing the hierarchical positioning of white students as the default standard of success.

### *The Ellisonian Self and the Invisibility of Students of Color*

The *Ellisonian self*, which emerges in CRT through the work of Taliaferro-Baszile (2009), captures the lived experience of invisibility faced by students of color in educational settings. This concept highlights how STEM policies, often developed without input from marginalized communities, fail to recognize the unique challenges faced by students of color, rendering them invisible within the broader educational discourse. For instance, science education reforms frequently adopt a one-size-fits-all approach, disregarding the cultural, social, and economic realities of students of color. Such policies neglect to address these students' specific needs, thereby marginalizing them within science classrooms and broader educational discourses. This invisibility within policy discussions and classroom practices not only marginalizes students of color but also perpetuates the broader societal view of them as subpersons, unworthy of equitable consideration in educational planning (Ellison & Morton, 1952).

**Answer to RQ 2: What role does racial commodification play in shaping the experiences of underrepresented students in STEM education, and how does it affect their academic and social outcomes?**

Racial commodification, a concept deeply rooted in Critical Race Theory (CRT), refers to the process by which the racial identities and bodies of underrepresented groups are exploited for institutional gain, often at the expense of the genuine needs and experiences of those individuals (Bridges, 2002). In the context of STEM education, racial commodification shapes the experiences of underrepresented students in significant ways, influencing both their academic and social outcomes. This process manifests in how educational institutions leverage the presence of these students to enhance diversity metrics and fulfill external mandates without substantively addressing the systemic barriers that hinder their academic success.

### *Commodification as a Superficial Diversity Metric*

Racial commodification in STEM education often occurs when institutions prioritize the representation of underrepresented students to project an image of diversity and inclusion, rather than implementing meaningful policies that support these students' academic and social development (Basile & Lopez, 2015). Schools and universities frequently highlight the enrollment of students of color in STEM programs to meet diversity quotas and attract funding or accolades, yet these superficial efforts fail to translate into real educational support. As a result, students of color become tokens within the system, celebrated for their presence but marginalized in their participation. This commodification contributes to an environment where students are valued primarily for their ability to fulfill institutional goals rather than as individuals with unique intellectual and cultural contributions.

### *Impacts on Academic Outcomes*

The commodification of underrepresented students in STEM has direct implications for their academic experiences and outcomes. By emphasizing numerical representation over substantive support, educational institutions often neglect the structural changes necessary to ensure these students' success. For instance, underrepresented students may be recruited into STEM programs without corresponding investments in culturally responsive teaching, mentoring, or resources tailored to their specific needs (Bonilla-Silva, 2015). The result is a mismatch between the institution's diversity rhetoric and the everyday realities of students, who may struggle with inadequate academic support, isolation, and a lack of belonging in STEM fields that have historically excluded them.

Moreover, the lack of genuine support can exacerbate academic challenges, such as under-preparedness for advanced coursework, limited access to high-quality instruction, and insufficient guidance on navigating the complexities of STEM disciplines. These barriers not only undermine students' academic performance but also reinforce harmful stereotypes that portray students of color as less capable or committed to STEM (Basile & Azevedo, 2022). This cycle of marginalization perpetuates racial hierarchies within STEM, as students of color are often blamed for their struggles rather than recognizing the systemic factors contributing to their difficulties.

### *Social and Psychological Effects of Commodification*

The effects of racial commodification extend beyond academic performance, significantly impacting the social and psychological well-being of underrepresented students. The process of being commodified and valued for their racial identity rather than their potential can lead to feelings of alienation and invisibility within educational spaces. Students may experience what Ellison and Morton (1952) describe as the "invisibility" of the marginalized self, where their presence is acknowledged, but their voices and experiences are systematically ignored. This invisibility fosters an environment in which underrepresented students are seen but not truly heard, leading to social isolation and a diminished sense of belonging.

The psychological burden of commodification can also manifest as internalized racial inferiority, where students begin to doubt their intellectual capabilities and question their place within STEM disciplines (Taliaferro-Baszile, 2009). The constant need to prove themselves in an environment that does not fully support them can result in increased stress, lower self-esteem, and a higher likelihood of disengagement from STEM pathways. This internal conflict reflects CRT's broader critique of how systemic racism not only constrains opportunities for students of color but also shapes their self-perception and agency within educational contexts.

### *Commodification and the Reproduction of Inequities*

Racial commodification within STEM education also serves to reproduce and reinforce broader societal inequities. By treating underrepresented students as tools for achieving diversity benchmarks rather than addressing the root causes of educational disparities, institutions contribute to the ongoing marginalization of these groups. This dynamic upholds existing power structures, allowing predominantly white institutions to benefit from the appearance of diversity without fundamentally altering the inequitable conditions that limit the success of students of color (Ladson-Billings, 1998).

For instance, when schools market their commitment to diversity to attract resources, they often fail to reinvest those resources in ways that directly benefit underrepresented students, such as through targeted academic support services or inclusive curriculum design. This disconnect between institutional priorities and student needs exemplifies the commodification process, whereby the labor and presence of students of color are exploited to enhance the institution's image without genuinely addressing racial inequities.

**Answer to RQ 3: How can STEM educators and institutions move beyond deficit-based narratives and adopt practices that affirm the personhood and intellectual capabilities of underrepresented students?**

Critical Race Theory (CRT) provides a powerful lens to examine and challenge the deficit-based narratives that persist within STEM education. Deficit-based narratives frame underrepresented students as lacking the skills, knowledge, or cultural capital necessary for success, often blaming students and their communities for systemic failures rather than addressing the structural inequities embedded in educational institutions (Yosso, 2005). Moving beyond these narratives requires STEM educators and institutions to adopt practices that affirm the personhood and intellectual capabilities of underrepresented students, recognizing their strengths, lived experiences, and inherent potential.

*Challenging Deficit Narratives Through Counter-Narratives*

CRT emphasizes the importance of counter-narratives as a tool to challenge dominant discourses that marginalize students of color. Counter-narratives center the voices and experiences of underrepresented students, highlighting their resilience, creativity, and intellectual contributions that are often overlooked in mainstream educational contexts (Solórzano & Yosso, 2001). By integrating these counter-narratives into STEM curricula and classroom practices, educators can disrupt deficit-based perspectives and affirm the personhood of students of color.

For instance, incorporating stories of Black, Indigenous, and other underrepresented scientists, engineers, and mathematicians into STEM education helps validate the presence and accomplishments of these communities within the field. This approach not only challenges the myth that STEM is a white-dominated space but also provides students with role models who reflect their own identities, reinforcing the message that they belong and can excel in STEM disciplines. Emphasizing the achievements and intellectual contributions of these figures disrupts the dominant narrative that only Euro-American individuals are capable knowers and innovators in STEM (Taliaferro-Baszile, 2009).

### *Adopting Asset-Based Pedagogies*

CRT advocates for asset-based pedagogies that recognize and leverage the cultural wealth of underrepresented students, viewing their backgrounds as assets rather than liabilities (Yosso, 2005). Asset-based approaches, such as culturally responsive teaching, validate students' lived experiences and integrate their cultural knowledge into the learning process. This pedagogical shift encourages educators to view underrepresented students not as passive recipients of knowledge but as active contributors with valuable perspectives that can enrich STEM learning.

Culturally responsive teaching practices include adapting instruction to reflect students' cultural contexts, using inclusive language, and designing learning activities that connect STEM concepts to students' everyday lives. For example, incorporating community-based projects that address local environmental or technological issues can make STEM learning more relevant and engaging for underrepresented students. These approaches acknowledge the intellectual capabilities of students of color, positioning them as co-creators of knowledge rather than subjects of remediation (Ladson-Billings & Tate, 1995).

### *Recognizing and Valuing Personhood in STEM Spaces*

CRT challenges the historical and ongoing marginalization of students of color in STEM by advocating for educational practices that affirm their personhood acknowledging their full humanity beyond stereotypes or preconceived notions of inferiority (Mills, 1997). This affirmation involves recognizing underrepresented students as complete individuals with complex identities, deserving of respect, dignity, and equitable opportunities in STEM education.

STEM educators and institutions can affirm personhood by creating inclusive and supportive learning environments that address the social and emotional needs of students. This might include fostering a classroom culture that values diverse perspectives, implementing mentorship programs with faculty who share similar cultural backgrounds, and providing safe spaces for students to discuss challenges related to race and identity. By valuing students' personhood, educators can combat the negative effects of racial stereotypes and build a foundation of trust and respect that enhances student engagement and achievement.

### *Moving Toward Equity-Driven Institutional Practices*

CRT emphasizes the need for systemic change within educational institutions to dismantle the structures that perpetuate racial inequities. For STEM education to move beyond deficit-based narratives, institutions must critically examine their policies, practices, and curricula to identify and address biases that disadvantage underrepresented students. This involves rethinking traditional assessment methods, admissions criteria, and placement practices that often serve as gatekeepers to STEM success for students of color.

Institutions can adopt equity-driven practices such as holistic admissions processes that consider a broader range of student achievements, equity-focused professional development for STEM educators, and the implementation of anti-racist policies that actively work to dismantle discriminatory barriers. These systemic changes can help shift the narrative from one of deficiency to one of equity, recognizing the value of diverse experiences and fostering an inclusive STEM community where all students are empowered to thrive (Ladson-Billings, 1998).

*Reframing Success and Redefining Excellence in STEM*

To affirm the intellectual capabilities of underrepresented students, redefining what constitutes success and excellence in STEM is essential. Traditional success metrics, such as standardized test scores or grades, often reflect and reinforce racial biases rather than true measures of a student's potential (Duncan, 2002). CRT calls for a reimagining of these metrics to include a broader understanding of achievement that values creativity, problem-solving skills, collaborative work, and community engagement.

Educators can employ alternative assessments, such as project-based learning, portfolios, or peer evaluations, which allow students to demonstrate their knowledge and skills in ways that align with their strengths. By redefining success, institutions can move away from rigid, exclusionary standards and create pathways that recognize and celebrate the diverse intellectual contributions of underrepresented students in STEM.

## Moving Beyond Gap Gazing: Toward Transformative Equity in STEM Education

To genuinely promote equity in STEM education, there is a need to move beyond the narrow focus of gap-gazing and address the systemic and cultural barriers that underlie educational inequities. This shift involves reimagining equity not as a means of simply raising the performance of marginalized students to match that of their peers but as a commitment to transforming educational environments to be inclusive, responsive, and just.

**Centering Asset-Based Frameworks**

One way to move beyond gap-gazing is by adopting asset-based frameworks, such as Yosso's (2005) Community Cultural Wealth model, which emphasizes the strengths and resources that marginalized students bring to the learning environment. By focusing on students' cultural, linguistic, and familial knowledge, educators can design STEM curricula that are relevant and engaging, fostering a sense of belonging and validating students' identities. This approach challenges the deficit-oriented narratives that dominate equity discourse and repositions students as capable and valuable contributors to STEM fields.

**Addressing Structural Inequities in STEM Education**

Equity initiatives must also confront the structural inequities that shape access and opportunities within STEM education. This includes examining discriminatory practices such as tracking, limited access to advanced coursework, and the underfunding of schools serving marginalized communities (Leonardo, 2009). Policies must be designed to dismantle these barriers by providing equitable resources, access to experienced teachers, and inclusive learning environments that reflect the diversity of student experiences.

**Culturally Responsive Pedagogy and Inclusive Teaching Practices**

Moving beyond gap-gazing requires a commitment to culturally responsive pedagogy that recognizes and integrates the cultural contexts of students' lives into STEM education. This includes teaching practices that value diverse ways of knowing, challenge dominant narratives, and create spaces where all students feel seen, respected, and capable of success (Gay, 2002). By creating culturally affirming STEM environments, educators can foster a more inclusive and equitable learning experience that goes beyond superficial measures of achievement.

**Engaging in Critical Self-Reflection and Systemic Change**

Finally, educators, policymakers, and institutions must engage in ongoing critical self-reflection to understand how their practices and policies contribute to or challenge systemic inequities. This involves questioning the traditional metrics of success, recognizing the limitations of current equity initiatives, and committing to transformative change that centers the needs and voices of marginalized communities. By prioritizing systemic change over superficial fixes, STEM education can move toward a more equitable and inclusive future.

**Conclusion**

In conclusion, this paper has critically examined how STEM education policies perpetuate racial hierarchies and inequities, using Critical Race Theory (CRT) as a framework to highlight the deep-rooted challenges faced by underrepresented students. The analysis reveals that seemingly neutral or inclusive policies often reinforce dominant norms that marginalize students of color, and that racial commodification reduces these students to institutional diversity symbols rather than affirming their personhood and intellectual potential. To counter these systemic barriers, STEM education must adopt policies and practices that recognize the unique experiences, strengths, and contributions of marginalized students.

**Moving Forward**

Moving forward, achieving equity in STEM education requires a fundamental shift in both policy design and institutional practices. Educators and policymakers must reject deficit-based narratives that view students of color through the lens of deficiency and instead foster environments that validate their cultural wealth and resilience. This shift necessitates the development of culturally responsive curricula, mentorship opportunities, and assessments that reflect the diverse learning styles and needs of all students. Institutions must also re-evaluate their diversity initiatives, ensuring that students are supported academically and socially rather than commodified for institutional branding.

The future implications of this work call for more than just policy reform; they demand a comprehensive transformation in how we think about and enact diversity, equity, and inclusion within STEM education. By centering the voices and experiences of marginalized communities and committing to the dismantling of racial hierarchies, the STEM field can move closer to a truly inclusive and equitable future. CRT provides a powerful lens to guide this process, pushing educators and institutions to critically examine their practices and challenge the status quo. In alignment with the NSF's goals for diversity and inclusion, future research and practice should focus on long-term strategies that not only address immediate inequities but also cultivate an environment where all students, regardless of race or background, can thrive in STEM disciplines.

44